\documentclass[article,
   amssymb,amsmath]{revtex4}

\usepackage[colorlinks=true,linkcolor=blue]{hyperref}%
\expandafter\ifx\csname package@font\endcsname\relax\else
 \expandafter\expandafter
 \expandafter\usepackage
 \expandafter\expandafter
 \expandafter{\csname package@font\endcsname}%
\fi
\usepackage{graphicx}
\usepackage{subfigure}
\usepackage{multirow}
\usepackage{footnote}
\usepackage{color}
\usepackage{epstopdf}
\usepackage{dcolumn}
\usepackage{bm}
	
\def\be{\begin{equation}}
\def\ee{\end{equation}}
\def\bea{\begin{eqnarray}}
\def\eea{\end{eqnarray}}
\def\na{\hat{N}_a}
\def\nazero{\tilde{N}_{a0}}
\def\naone{\tilde{N}_{a1}}

\begin{document}

\title{Dynamic Landau Theory for Supramolecular Self-Assembly}

\author{Nitin S. Tiwari}
\email[Author to whom correspondence should be addressed. ]{n.tiwari@tue.nl}
\affiliation{Department of Applied Physics, Eindhoven University of Technology,
P.O. Box 513, 5600 MB Eindhoven, The Netherlands}
\author{Koen Merkus}
\affiliation{Department of Applied Physics, Eindhoven University of Technology,
P.O. Box 513, 5600 MB Eindhoven, The Netherlands}
  \author{Paul van der Schoot}
\affiliation{Department of Applied Physics, Eindhoven
University of Technology, P.O. Box 513, 5600 MB Eindhoven, The Netherlands}
\affiliation{Institute for Theoretical Physics, Utrecht University,
Leuvenlaan 4, 3584 CE Utrecht, The Netherlands}

\date{\today}

\begin{abstract}
Although pathway-specific kinetic theories are fundamentally important to describe and understand reversible polymerisation kinetics, they come in principle at a cost of having a large number of system-specific parameters. Here, we construct a dynamical Landau theory to describe the kinetics of activated linear supramolecular self-assembly, which drastically reduces the number of parameters and still describes most of the interesting and generic behavior of the system in hand. This phenomenological approach hinges on the fact that if nucleated, the polymerisation transition resembles a phase transition. We are able to describe hysteresis, overshooting, undershooting and the existence of a lag time before polymerisation takes off, and pinpoint the conditions required for observing these types of phenomenon in the assembly and disassembly kinetics. We argue that the phenomenological kinetic parameter in our theory is a pathway controller, i.e., it controls the relative weights of the molecular pathways through which self-assembly takes place.

\end{abstract}

\pacs{Valid PACS appear here}
\maketitle


\section{\label{sec:level1} Introduction}
Supramolecular polymerisation, $\beta-$amyloid fibril formation, actin and microtubule polymerisation all have two features in common: (i) some form of activation and (ii) reversible elongation.\cite{tomdegreef, paul3, oosawa, lomakin} Activation can happen in many different ways, the most important being conformational switching \cite{bibal} and a minimum number of monomers coming together before elongation can take place (nucleation).\cite{oosawa, greer} Generally, the activation constant for the above-mentioned systems is very small compared to the elongation constant, giving rise to very sharp polymerisation transition as a function of, e.g., the temperature, concentration, acidity and so on. This makes the polymerisation transition reminiscent of a phase transition.\cite{tobolsky, wheeler} Hence, if we keep aside details of the actual activation mechanism and other system-specific details, we can hope to understand the universal and the most interesting behavior of many such systems by relying on notions from the statistical mechanics of the phase transitions.\cite{paul3, wheeler}

Even though our understanding of the thermodynamics of supramolecular polymers and the role of activation or nucleation,\cite{nucleation1, nucleation2} solvent,\cite{solvent1, solvent2, solvent3} conformational \cite{conformation1, conformation2} and compositional disorder \cite{composition1, composition2} has made great advances, much less is known about the \mbox{\it{kinetics}} underlying reversible polymerisation processes.\cite{paul3} The most extensively studied kinetic reaction rate models were initially set up to describe the equilibration of the length distribution of worm-like surfactant micelles in response to a temperature jump, and stress relaxation under shear flow via the dynamical breakdown and re-growth of polymeric assemblies.\cite{cates1} Four kinetic pathways have been identified in this context: (i) end evaporation and addition,\cite{cates1} (ii) scission and recombination,\cite{cates2} (iii) end-interchange,\cite{cates3} and (iv) bond-interchange.\cite{cates4} In principle, one should consider not a single but a hybrid of pathways,\cite{cates5} but this is rarely done.\cite{cohen1}

The reaction rate or master equations are invariably highly non-linear integro-differential equations, some of which elude exact analytical solutions even in linearised form.\cite{cates1} Only in some limiting cases asymptotic analytical results have been obtained for the temporal evolution of the length distribution within the end-evaporation-and-addition and scission-recombination kinetics applying the rate-equation approach outside of the linearised regime.\cite{semenov1, semenov2, paul4} Not surprisingly, it is tempting to obtain closed-form equations for first few moments only, presumably requiring the assumption that the shape of the probability distribution function does not change with time. This in turn requires pathways to be tuned in such a way that it may not be possible to obtain correct thermodynamic equilibrium (e.g., by assuming irreversible scission).\cite{cohen1, cohen2, cohen3}

In this paper we present as an alternative to the above-mentioned approaches, a phenomenological dynamical Landau theory for activated reversible polymerisation processes that makes use of what is known about the kinetics of phase transitions. The advantage of this method is that it allows for straightforward coupling of the equilibrium  polymerisation to other kinds of macroscopic phase transition, such as phase separation and the spontaneous alignment in liquid crystalline states \cite{intransition1, intransition2, intransition3} as well as to flow fields.\cite{shear1, shear2} The coupling is straightfoward because all the phenomena (polymerisation, phase ordering of various kind) can be treated on an equal footing in terms of appropriate order parameters. Here, we focus on studying the dynamics of  polymerisation in the absence of any symmetry breaking and flow fields, leaving that for future work. Our theory is able to describe kinetic phenomena like hysteresis, overshooting, undershooting and the existence of a lag time observed experimentally, yet has only a single kinetic parameter. We argue that this parameter controls in some sense the relevant weights of different molecular pathways implicit in the theory. We find that our theoretical predictions are in qualitative agreement  with experimental observations on the assembly kinetics of $\beta-$amyloids, at least if we renormalize the theory with an appropriate time scale.

The remainder of this article is organized as follows. In section II, we construct a thermodynamically consistent Landau free energy involving two relevant non-conserved order parameters, representing two moments of the full distribution function, and then use that to construct our dynamical equations.\cite{hohenberg} The two moments we focus attention on are the polymer fraction and degree of polymerisation. We ignore any spatial variation and presume the system is well mixed at all times. Our two nonlinear differential equations describing the temporal evolution of the two order parameters in essence depend on only two dimensionless groups. One, the \mbox{\it{mass action variable}} describes the strength of the thermodynamic driving force towards polymerisation and the other, \mbox{\it{the pathway controller variable}}, selects the relaxation pathway.
In section III, we solve the linearised version of these equations and investigate transient phenomena known as ``overshooting'' and ``undershooting''. The former has been observed in nucleated self-assembling polymer solutions of tobacco mosaic virus (TMV) coat protein \cite{schuster} and of actin.\cite{brooks} We analyse the delayed response before assembly takes off, an observable studied extensively in different types of system, in section IV. There, we solve our nonlinear evolution equation using the method of ``Matched Asymptotic Expansion'' (MAE) and obtain, in analytical form, the lag time as a function of the relevant mass action variable and the initial conditions. In section V, we provide an interpretation for our pathway control variable. We also discuss the phenomenon of temporal hysteresis and match our analytical results with experimental data for $\beta-$amyloid fibril assembly and obtain excellent agreement if we allow for an offset. We end this paper with a summary and conclusion in section VI.

\section{Landau Free Energy Function and Dynamical Equations}
Our problem of interest is activated polymerisation in dilute yet well-mixed solution. This implies that we consider two basic components: assembly active polymers and assembly inactive monomers. Inactive monomers have to acquire a high-energy, activated state to be able to polymerize. Activated monomers convert inactive into active ones upon binding. If the latter does not require any free energy input, this is known as ``autosteric'' or ``auto-catalytic'' binding.\cite{paul3, paul5} If it does require a free energy input, we have conventional activated polymerisation.\cite{aggelli,dudowicz} Both types of polymerisation turn out to obey the same statistics, i.e., the mass action models that describe these statistics are equivalent.\cite{paul3}

We arbitrarily choose the autosteric model. In it, the free-energy difference between active and inactive states is $\Delta f_a \geq 0$, and the free-energy gain upon bonding is $\Delta f_e \leq 0$. Let $\phi$ be the overall concentration (mass fraction) of monomers in the solution. If we now invoke the law of mass action and assume that the free monomers, the dimers, the trimers, etc., do not mutually interact, we obtain for the active polymers an exponential distribution with an average degree of polymerisation that we denote $\bar{N}_a$. A fraction $f$ of the material is in the polymerised, i.e., active state. The overall mean aggregation number, including active and inactive species, we denote $\bar{N}$. Under conditions of thermodynamic equilibrium, one can show that  \cite{paul5}
\bea
f = \frac{\bar{N}_a (\bar{N}_a -1) K_a}{X},
\eea
where $\bar{N}_a$ obeys an equation of state given by
\bea
X=1-\bar{N}_a^{-1}+K_a \bar{N}_a (\bar{N}_a-1).
\eea
Here, $X \equiv \phi \exp{\left( -\Delta f_e / k_B T \right)}$ is our mass-action variable and $K_a \equiv \exp{\left( -\Delta f_a / k_B T \right)}$ the activation constant, where, $k_B T$ denotes the thermal energy with $k_B$ Boltzmann's constant and $T$ the absolute temperature. Note that the mean degree of polymerisation averaged over active and inactive species obeys $\bar{N} = \left( 1+K_a \bar{N}_a^2 \right) / \left(1+K_a \bar{N}_a \right)$.

From  Eqn. (1) and (2), we deduce that $f=(X-1+\bar{N}_a^{-1})/X$. If we demand that $K_a \rightarrow 0$ and $X \geq 1+K_a$, then $f \sim 1 - X^{-1} + X^{-1} \sqrt{K_a (X-1)^{-1}} \sim 1-X^{-1}+O(K_a^{1/2})$. So, indeed, the polymerisation transition becomes infinitely sharp in the limit $K_a \rightarrow 0$, with
\begin{equation}
  f \sim \begin{cases}
    0 & \text{for $X \leq 1$},\\
    1-X^{-1} & \text{for $X \geq 1$}.
  \end{cases}
\end{equation}
It is straightforward to show that in this limit the heat capacity exhibits a jump at the polymerisation point $X=1$, as to be expected from mean-field arguments.\cite{paul6, tomdegreef} The limit $K_a \rightarrow 0$ is sensible because experimental values are typically $10^{-2}-10^{-5}$.\cite{paul_charley, nucleation2, solvent1}

Now that we have convinced ourselves that the polymerisation transition resembles a phase transition,\cite{paul6, wheeler} we may attempt to describe it by constructing a Landau free energy that will become a starting point of a dynamical theory. We recall that there are two types of distribution: an exponential size distribution of active polymers and a distribution over active and inactive material. This implies that we should be able to describe the thermodynamics of activated polymerisation with only two order parameters, one representing the fraction of polymerised material $f$ and the other describing the mean aggregation number of active material $\bar{N}_a$.
From Eqs. (1) and (2) we conclude that whilst $f$ is critical in the limit $K_a \rightarrow 0$, $\bar{N}_a$ is not. In this limit $f$ is non-zero only if $X > 1$ yet $\bar{N}_a > 1$ for all $X>0$. Indeed, from Eqn. (2) we find that at $X=1$, $\bar{N}_a = K_a^{-1/3} \gg 1$ if $K_a \ll 1$. Hence, only $f$ exhibits a sharp transition at the polymerisation point.\cite{paul3}

Parenthetically we note that for $\bar{N}_a \gg 1$, Eq. (1) suggests that $fX \sim K_a \bar{N}^2_a$ and it seems therefore that as $f$ becomes critical, so does $\bar{N}_a$. However, because we are interested in the limit $K_a \rightarrow 0$, the product $K_a \bar{N}^2_a$ going to zero does not mean that $\bar{N}_a$ also goes to zero or even to unity as we have just seen. Hence, it makes sense not to use $\bar{N}_a$ as order parameter but instead a quantity proportional to $K_a \bar{N}^2_a$. In this case both order parameters are zero below the critical point $X=1$, and non-zero and finite above it even in the limit $K_a \rightarrow 0$.

It is important to point out in this context that our statistical mechanical model is identical to the Tobolsky-Eisenberg model for (spontaneous) sulfur polymerization, which is an example of activated equilibrium polymerization.\cite{tobolsky} Wheeler and Pfeuty\cite{wheeler} have shown that a magnetic spin lattice model, the so-called n-vector model, becomes equivalent to the Tobolsky-Eisenberg theory in the limit $n \rightarrow 0$.
In that model, the spin-spin coupling constant corresponds to our elongation constant $K_e$, whereas the magnetic field strength maps onto the activation constant $K_a$. Wheeler and Pfeuty show that in their prescription the number concentration of polymers depends linearly on the external field. Hence, in the limit of zero magnetic field, the number concentration of polymers tends to zero. A vanishing number concentration of polymer chains has to accommodate a finite polymerized mass, giving rise to a diverging mean degree of polymerization above the critical point. This confirms that it is not sensible to use $\bar{N}_a$ as an order parameter in the limit $K_a \rightarrow 0$.

It follows that a sensible Landau free energy describing nucleated polymerisation must involve two coupled order parameters. In a state of thermodynamic equilibrium, this free energy should retrieve Eqn. (1) and (2) or (1) and (3), which are equivalent in the appropriate limit $K_a \rightarrow 0$. To stay within the philosophy of Landau theory we opt for the latter, because it allows us to invoke the relevant control variable, $X$, and let it play a role similar to temperature. The mass action variable $X=\phi \exp{\left( -\Delta f_e / k_B T \right)}$ depends on the solution conditions, that is, the concentration $\phi$, the temperature $T$ and, depending on the type of system, on the solvent, the acidity and the salinity that (apart from temperature) control the binding free energy $\Delta f_e$.

As we have seen, the polymer fraction $0 \leq f \leq 1$ shows a sharp continuous transition, which is a signature of a second order phase transition, requiring a Landau free energy function consisting of even powers of the corresponding order parameter. To satisfy two essential properties of the theory, i.e., a second-order-like transition and non-negativity of polymer fraction $f$, we select this order parameter to be $\sqrt{f}$ instead of $f$. Hence, we denote our first order parameter as $S_1=\sqrt{f}$, giving us the first two terms in our free energy (per particle) as $-(X_p^{-1} -X^{-1}) S_1^2 + S_1^4$, in accordance with the requirements of the free energy for the second order phase transition, and showing a transition at $X=X_p \equiv 1$, $X_p$. $X_p$ is transition point of the control variable $X$, even though minimizing this free energy does not quite produce Eqn. (3) yet as it is off by a factor of two. This, we fix below.

Apart from the critical order parameter $S_1$ that is related to the fraction of active material $f$, we have to introduce another order parameter, $S_2$, that somehow describes the degree of polymerisation of the active material, $\bar{N}_a$, and that is a proper order parameter. As already suggested, an in this context natural order parameter would be $S_2=\bar{N}_a (\bar{N}_a -1) K_a/X$, which is also critical but which in our model is enslaved by $S_1$. See Eq. (1).

This suggests a contribution to the free energy density a term proportional to $S_2^2$, and a coupling term that establishes the enslavement of $S_2$ to $S_1$. We have to keep in mind that in equilibrium we have to obey eqn. (1). This we manage by adding to our free energy the terms $(1/2) S_2^2 - S_1^2 S_2$. The introduction of the coupling term $S_1^2 S_2$ ensures that we automatically obey Eqn. (1) \textit{and} (3) in equilibrium, that is, if we minimize the free energy with respect to $S_1$ and $S_2$.

In conclusion, we obtain the following free energy per solute molecule $F$ that obeys,
\bea
\frac{F}{k_B T} = -(1 -X^{-1}) S_1^2 + S_1^4 + \frac{1}{2} S_2^2 - S_1^2 S_2,
\eea
and that meets all requirements. Note that as usual $k_B T$ denotes the thermal energy. As another thermodynamic consistency check, we calculate the specific heat using this free energy. It turns out to be consistent with that obtained invoking the law of mass action (or the equivalent microscopic thermodynamic theory).\cite{paul6} We refer to Appendix A for details.

Now that we have defined our free energy landscape, we can build on that our dynamical theory. Using the appropriate description for non-conserved order parameters (model A),\cite{hohenberg} we set
\begin{align}
\frac{\partial S_1}{\partial t} &= - \Gamma_1 \frac{\partial F/k_B T}{\partial S_1} \quad  \text{and} \quad
\frac{\partial S_2}{\partial t} = - \Gamma_2 \frac{\partial F/k_B T}{\partial S_2},
\end{align}
with $\Gamma_1$ and $\Gamma_2$ phenomenological relaxation rates for our order parameters $S_1$ and $S_2$. In line with common practice, and in the absence of a clear microscopic interpretation, we presume these relaxation rates to be independent of the order parameters. As we shall see, this simplification will prove sufficient to produce all known dynamical behaviours of supramolecular assembly, including a lag phase, overshoots and hysteresis.

The resulting dynamical equations now read
\begin{align}
\frac{\partial S_1}{\partial \tau} &= 2 (1 - X^{-1}) S_1 + 2 S_1 S_2 - 4 S^{3}_1, \\
\frac{\partial S_2}{\partial \tau} &= \gamma (S^{2}_1 - S_2),
\end{align}
where we introduced a dimensionless time, $\tau \equiv \Gamma_1 t$, and a the ratio of relaxation rates $\gamma \equiv \Gamma_1/\Gamma_2$ that will prove to be our kinetic pathway controller. In principle, this could point towards a microscopic interpretation of the kinetic parameters and a possible evaluation of whether, and if so, how, they should depend on the order parameters. We leave this for future work.

Superficially, this set of coupled dynamical equations looks very simple. However, their highly non-linear character heralds not only a complex dynamical behavior but also difficulty in dealing with them analytically. Their non-linear character can be reduced somewhat by a simple transformation: $f \equiv S_1^2$ and $\na \equiv K_a \bar{N}_a (\bar{N}_a - 1 ) \equiv S_2 X$. Here, $f$ is as before the fraction of polymerised material, and $\na$ is a measure for the degree of polymerisation of the active material that from now on we call the renormalised active degree of polymerisation or renormalised mean polymer length. Inserting this into Eqn. (6) and (7), we obtain
\bea
\frac{\partial f(\tau)}{\partial \tau} &=& 4 (1 - \frac{1}{X}) f(\tau) + 4 f(\tau) \frac{\na (\tau)}{X} - 8 f(\tau)^2, \\
\frac{\partial \na (\tau)}{\partial \tau} &=& \gamma f(\tau) - \frac{\gamma}{X} \na (\tau).
\eea
Note that if we solve the above set of dynamical equations with the initial condition $f(0)=0$, they will not evolve towards the correct equilibrium point, i.e., the minimum of the free energy. This is to be expected, as our theory is a dynamical mean field theory, if the starting point is a maximum free energy state, the order parameters will not evolve in the absence of fluctuations. In the present study, we choose not to include noise and hence focus on ``seeded polymerization''.
Also Notice that the highest order non-linear term in Eqn. (8) is now quadratic rather than cubic as is the case in Eqn. (6), and the quadratic term in Eqn. (7) becomes linear in Eqn. (9). This makes a linear analysis more accurate. As we shall show in the next section, even at the linearised level our governing equations give rise to the interesting transient phenomena of overshooting and undershooting.

\section{Overshoot and Undershoot}
An exact analytical solution of our set of governing dynamical equations, Eqn. (8) and (9), has evaded us. A numerical evaluation of Eqn. (8) and (9) in Fig. (1) shows that if we perturb the fraction of polymerised material and/or their mean aggregation number away from their equilibrium values, these quantities do not necessarily relax in a simple exponential fashion, but can exhibit a  transient response of non-monotonic growth or decay before approaching equilibrium. There are two types of non-monotonic relaxation that we call \mbox{\it{overshooting}} and \mbox{\it{undershooting}}.\cite{schuster} What this means precisely, will become clear below. Note that overshooting of the polymerised fraction has been observed in actin assembly,\cite{brooks} and that of the active degree of polymerisation in TMV coat protein assembly.\cite{schuster} We have not been able to find examples of undershooting; most experimental studies focus on assembly rather than disassembly.

Interestingly, overshooting and undershooting present themselves to us even in a linearised version of the governing equations, which we can solve exactly. The analytical solution allows us to demarcate different kinetic regimes dominated by overshooting, undershooting and simple exponential relaxation for both quantities $f$ and $\na$. To linearize eqns. (8) and (9) we write $f(\tau) = f(\infty) + \delta f(\tau)$ and $\na (\tau) = \na (\infty) + \delta \na (\tau)$. Here, $f(\infty)$ and $\na (\infty)$ are the respective equilibrium values for time $\tau$ going to infinity, and $\delta f(\tau)$ and $\delta \na (\tau)$ are the perturbations away from equilibrium, starting at zero time, $\tau=0$. The solutions to our linearised version of our dynamical equations read
\bea
\delta f(\tau) &=& A_1 e^{-\lambda_1 \tau} + B_1 e^{-\lambda_2 \tau},
\eea
and
\bea
\delta \na (\tau) &=& A_2 e^{-\lambda_1 \tau} + B_2 e^{-\lambda_2 \tau}.
\eea
These solutions are linear combination of eigenmodes with principal relaxation rates (the eigenvalues of the dynamical matrix),
\bea
\lambda_1 &=& -\frac{8-8X-\gamma - \kappa}{2 X},
\eea
and
\bea
\lambda_2 &=& -\frac{8-8X-\gamma + \kappa}{2 X}.
\eea
The amplitudes $A_1, B_1$ and $A_2, B_2$ of the corresponding eigenmodes are,
\bea
\hspace{-10pt} A_1 &=& \frac{\delta f(0) \left(\kappa+8X-\gamma-8\right)-8 (1-X^{-1}) \delta \na(0)}{2 \kappa}, \nonumber \\
\hspace{-10pt} B_1 &=& \frac{\delta f(0) \left(\kappa-8X+\gamma+8\right) + 8 (1-X^{-1}) \delta \na(0)}{2 \kappa},
\eea
and
\bea
A_2 &=& \frac{\left(\kappa-8X+\gamma+8\right) \delta \na(0)-2 \gamma  \delta f(0) X}{2 \kappa}, \nonumber \\
B_2 &=& \frac{\left(\kappa+8X-\gamma-8\right) \delta \na(0)+2 \gamma  \delta f(0) X}{2 \kappa}.
\eea
Here, $\kappa \equiv \sqrt{\gamma ^2+64 (X-1)^2}$. Notice that these solutions are meaningful for only $X>1$, because we cannot perform a perturbation theory for $X \leq 1$ as the equilibrium solution in this case is $f(\infty)=0$. Also note that the other limit $X \rightarrow \infty$ is not meaningful either, as in the linearised version of our dynamical equations, (8) and (9), the equilibrium conditions are, $\delta f (\infty) = \na(\infty) / 2X$ and $\delta f (\infty) = \na(\infty) / X$. Both these equilibrium conditions are true if and only if $\delta f (\infty) = \delta \na(\infty) = 0$, for any finite value of $X$. But in the limit $X \rightarrow \infty$, the equilibrium solution, $(\delta f (\infty), \delta \na(\infty)) = (0,0)$, becomes unstable.

\begin{figure*}[ht!b]
        \centering
        \begin{subfigure}
            \centering
            \hspace{-13pt} \includegraphics[width=6.5in]{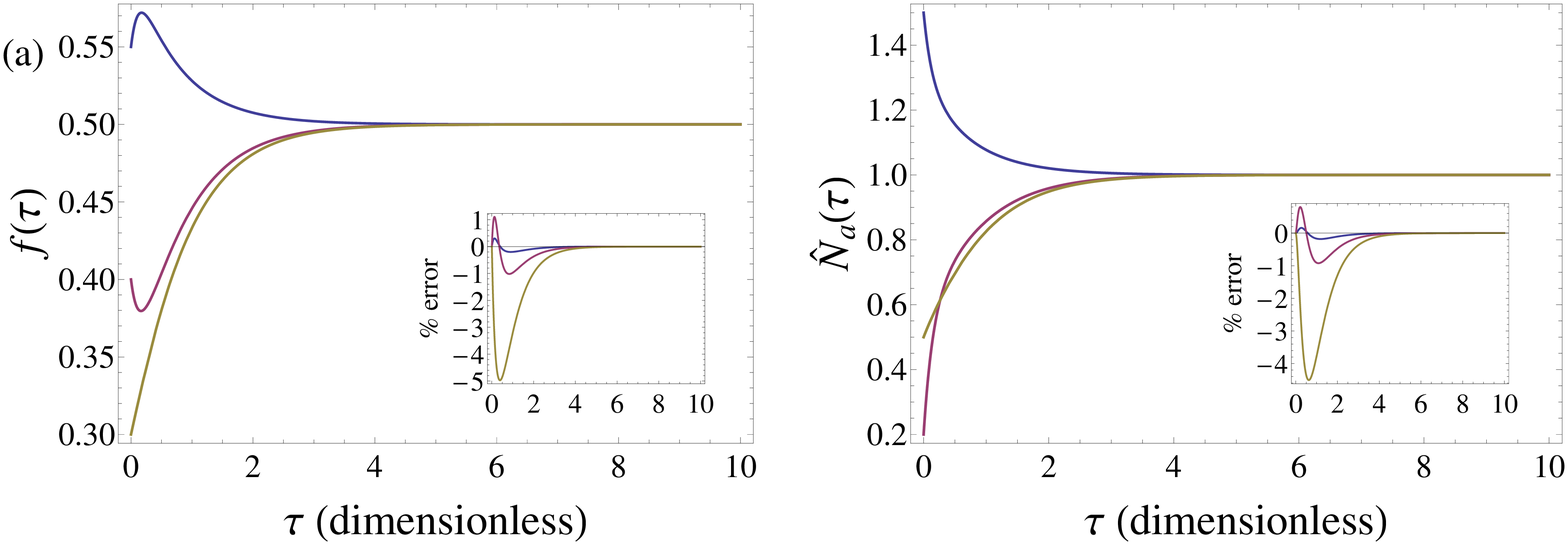}
        \end{subfigure}
        \hfill
        \begin{subfigure}
            \centering
            \includegraphics[width=6.5in]{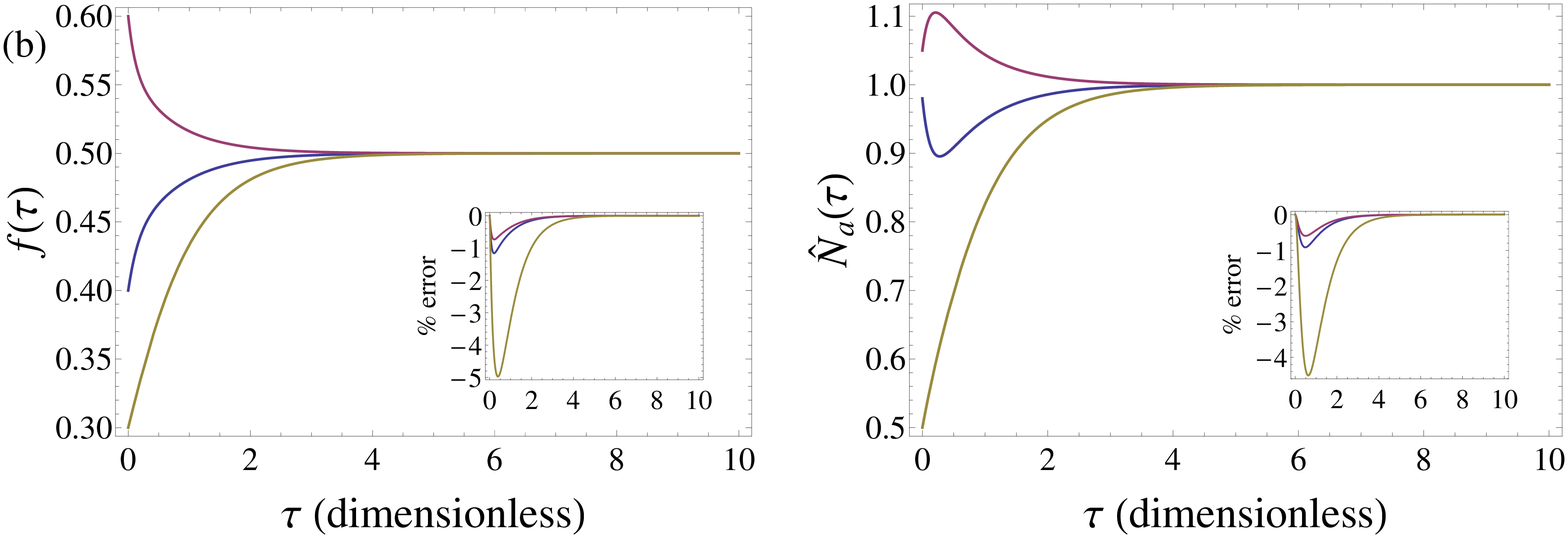}
            \begin{minipage}{.1cm}
            \vfill
            \end{minipage}
        \end{subfigure}
        \vspace{-15pt}
        \caption{ \label{fig:epsart} (a) Polymer fraction $f$ \mbox{\it{vs}} dimensionless time $\tau$ (left), and the quantity $\na=K_a \bar{N}_a^2$, with $K_a$ the activation constant and $\bar{N}_a$  the mean degree of polymerisation of the active matter, as a function of dimensionless time $\tau$ (right). Curves shown are for the initial conditions $f(0)=0.55,0.4,0.3 $ and $ \na(0)=1.5,0.2,0.5$, respectively, indicated in blue, purple and yellow. (b) Similar curves for initial conditions $f(0)=0.4,0.6,0.3$ and $\na(0)=0.98,1.05,0.5$, again blue, purple and yellow. For all figures we set the mass action variable at the value of $X=2$ and the pathway controller at a value of $\gamma=10$. See the main text for an explanation of these parameters. Given in the insets are the percentage differences (errors) of $f$ and $\na$ between linearised theory and numerical calculation.}
\end{figure*}

Fig. (1) shows that both numerical and linear solutions point at the existence of overshooting, undershooting and exponential monotonic relaxation. We never find overshooting and undershooting in both observables. What kind of response and in what variable we find the different types of response depends entirely on the initial conditions, i.e., on the values that our two order parameters have at time zero. The demarcation between different types of behavior we can evaluate using the linearised theory. For this purpose, we define the (dimensionless) time of transient response, $\tau^{\text{tr}}$, as the time at which the quantity showing a non-monotonic transient response is at its extremum. From Eqn. (10) and (11) we deduce that
\bea
\tau_f^{\text{tr}} &=& \frac{1}{\lambda_1 - \lambda_2} \ln \left( -\frac{A_1 \lambda_1}{B_1 \lambda_2} \right),
\eea
and
\bea
\tau_{\na}^{\text{tr}} &=& \frac{1}{\lambda_1 - \lambda_2} \ln \left( -\frac{A_2 \lambda_1}{B_2 \lambda_2} \right),
\eea
where $\tau_f^{\text{tr}}$ and $\tau_{\na}^{\text{tr}}$ are the transient response times for the quantities $f$ and $\na$, respectively. For a non-monotonic transient response to exist, these transient times must be real and positive, which can be evaluated using Eqn. (12) through (16). The transient times cannot be zero for the solutions obtained from a linear analysis because of the absence of any sigmoidal response at that level of approximation. As overshooting and undershooting can occur in both $f$ and $\na$, but never in both, this gives rise to four sets of initial conditions for each type of the transient response. We tabulate these in Table I, for positive and negative perturbation around the equilibrium values, and notice that the demarcation of the various regimes depend on both $X$ and $\gamma$, so on thermodynamics and kinetics.

\begin{table*}
\vspace{20pt}
 \centering
 \caption{Initial condition regimes showing transient response or monotonic response for the polymer fraction $f$ and $\na \equiv K_a \bar{N}_a^2$, where $K_a$ and $\bar{N}_a$ are the activation constant and the mean degree of polymerisation. The quantities, $\delta f$ and $\delta \bar{N}_a$, indicate perturbation around equilibrium at time zero.}
 \begin{tabular}{c | c | c|}
 \cline{2-3}
 \quad & $\delta f(0)>0$ & $\delta f(0)<0$ \\
 \cline{1-3}
 \multicolumn{3}{|c|}{transient response in $f$} \\
 \cline{1-3}
 \multicolumn{1}{|c|} {$\delta \na(0) > 0$} & $\delta \na(0) \ge 2 X \delta f(0)$ & $\delta \na(0) > \kappa_{-}$ \\
 \cline{1-3}
 \multicolumn{1}{|c|} {$\delta \na(0) < 0$} & $\delta \na(0) < \kappa_{+}$ \footnote{$\kappa_{\pm} \equiv \frac{1}{8} X \delta f(0) \left(\pm\sqrt{\frac{\gamma^2}{(X-1)^2}+64}+\frac{\gamma }{1-X}+8\right)$} & $\delta \na(0) \le 2 X \delta f(0)$ \\
 \cline{1-3}
 \multicolumn{3}{|c|}{transient response in $\na$} \\
 \cline{1-3}
 \multicolumn{1}{|c|} {$\delta \na(0) > 0$} & $\delta \na(0) \le X \delta f(0)$ & $\delta \na(0) < \kappa_{-}$ \\
 \cline{1-3}
 \multicolumn{1}{|c|} {$\delta \na(0) < 0$} & $\delta \na(0) > \kappa_{+}$ & $\delta \na(0) \ge X \delta f(0)$ \\
 \cline{1-3}
 \multicolumn{3}{|c|}{monotonic response in both $f$ and $\na$} \\
 \cline{1-3}
 \multicolumn{1}{|c|} {$\delta \na(0) > 0$} & $ \qquad X \delta f(0) < \delta \na(0) < 2 X \delta f(0) \qquad$ & no monotonic response \\
 \cline{1-3}
 \multicolumn{1}{|c|} {$\delta \na(0) < 0$} & no monotonic response & $ \qquad 2 X \delta f(0) < \delta \na(0) < X \delta f(0) \qquad $ \\
 \hline
 \end{tabular}
\end{table*}

Table I only tells us whether or not there is a transient response in $f$ or $\na$, dependent on the initial conditions. It does not tell us the nature of the transient response. In principle, we can deduce this by calculating the curvature of the quantity showing transient response at the point of the extremum, that is, by calculating the second derivative of $f$ and $\na$ with respect to time. Conditions that we derive from the second-derivative test turn out to be extremely complicated expressions, and hence we use a simpler method to find what kind of transient response our order parameters show. It hinges on evaluating the slope of $f$ and $\na$ at zero time and connecting that to the expected response from Table I. For instance, if conditions are such that $f$ shows transient behavior and if the slope at zero time is negative (positive) then we know that we are dealing with undershooting (overshooting).

Fig. (2) shows how the initial conditions dictate whether we have overshooting, undershooting or monotonic relaxation. Although this ``phase diagram'' is inferred from a linear analysis, it gives us information as to when overshooting or undershooting will occur for a specified set of parameters. The nonlinear terms in our dynamical equations modifies the extent of overshoot or undershoot and magnitude of transient time $\tau_{tr}$, and also slightly deform the boundaries separating various transient responses in Fig. (2) away from the equilibrium point. The phase diagram shows that one cannot, as already advertised, have non-monotonic response in both quantities $f$ and $\na$, a fact that can be inferred from Table I also. Indeed, Table I is exhaustive in describing all regimes of initial conditions that are mutually exclusive among the three different kinds of response.

\begin{figure}[ht!b]
        \centering
        \begin{subfigure}
            \centering
            \includegraphics[width=3.3in]{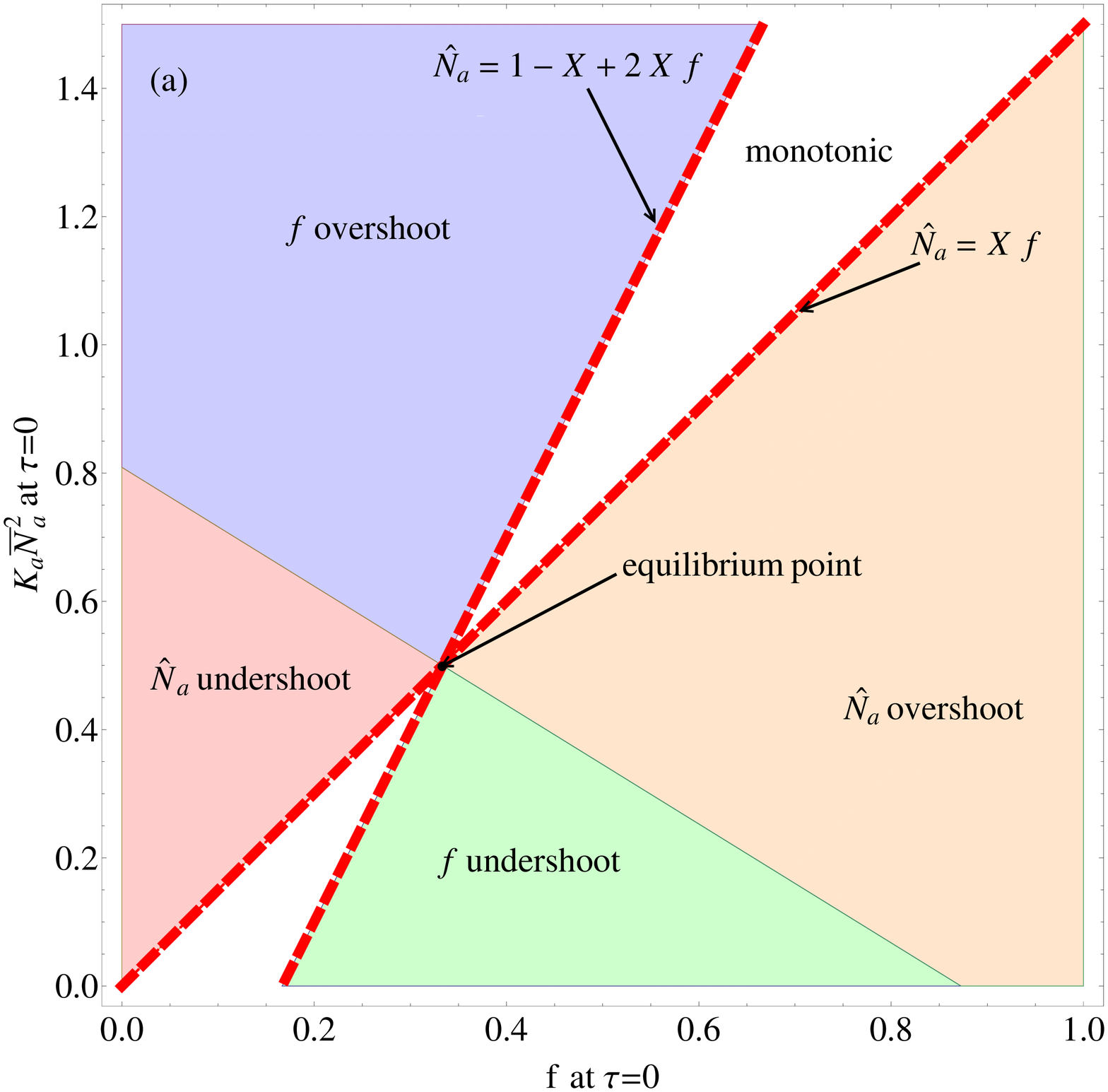}
        \end{subfigure}
        \hfill
        \begin{subfigure}
            \centering
            \includegraphics[width=3.3in]{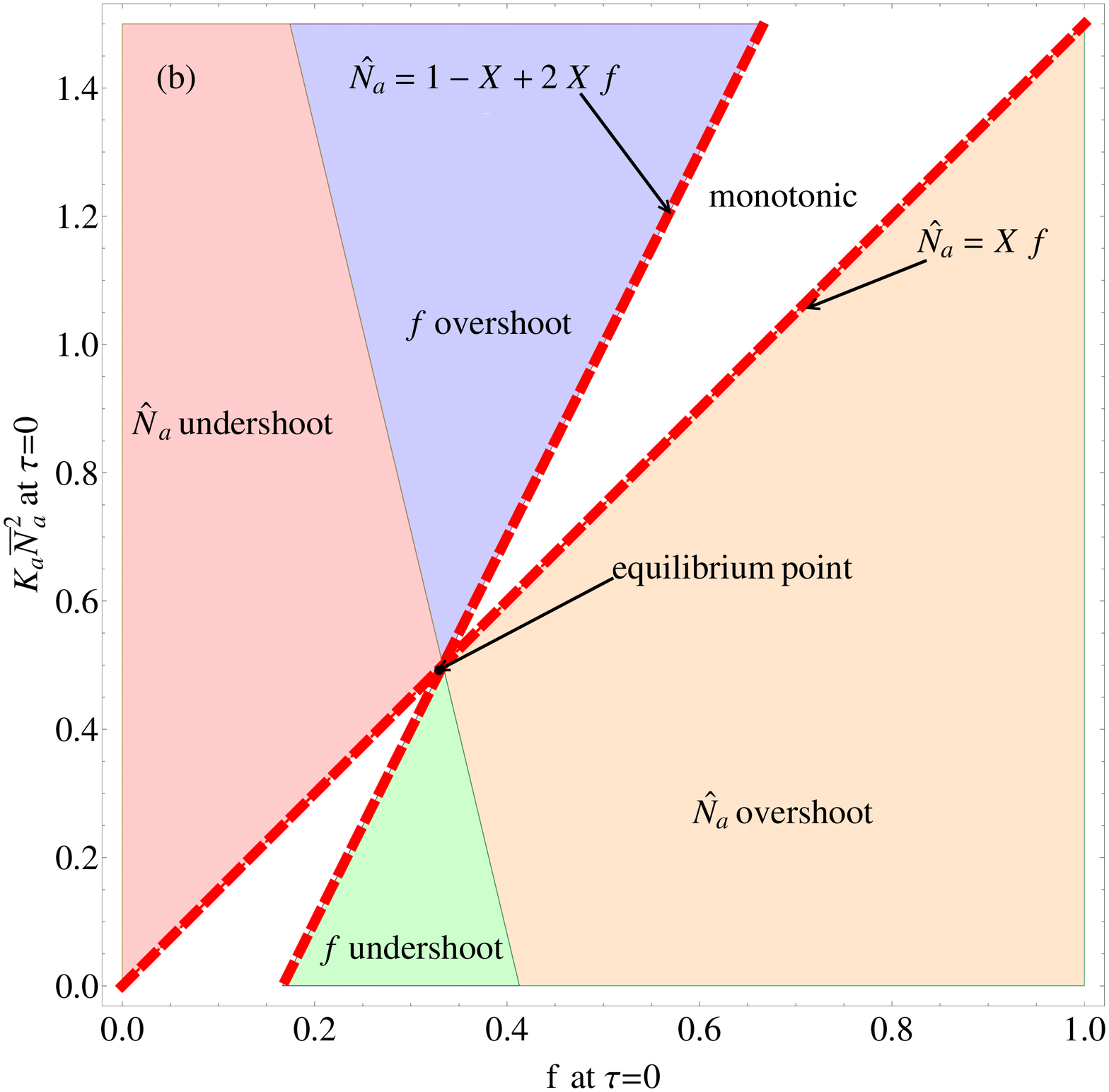}
            \begin{minipage}{.1cm}
            \vfill
            \end{minipage}
        \end{subfigure}
        \vspace{-15pt}
        \caption{``Phase diagram" showing the linear response of polymer fraction $f$ and $\na \equiv K_a \bar{N}_a^2$ as a function of their initial conditions. Different regions correspond to different kinds of transient kinetic responses. a) $X=1.5$ and $\gamma=2$, b) $X=1.5$ and $\gamma=10$}
\end{figure}

We can qualitatively understand the transient response of our order parameters from the dynamical equations, (8) and (9), and the equalities obtained from minimizing the free energy yielding the equalities, 1) $\na=1-X+2Xf$ and 2) $\na=Xf$, which we may call local equilibrium conditions. As shown in Fig. 2, if the initial conditions are such that the system starts in the region between the local equilibrium conditions then no transient response is observed. If we start with initial conditions away from the equilibrium point, the system first transiently evolves towards either condition 1 or 2, depending on the initial conditions and once that has been accomplished, both the order parameters simultaneously decay to satisfy Eqn. (3). In the process of $f$ following $\na$, or vice versa depending on the initial conditions and the value of pathway controller, our order parameters show overshoot, undershoot or monotonic response. This implies that the only transition that is sensitive to the value of $\gamma$, is that between undershoot and overshoot.

Obviously, it would be very useful to have a physical interpretation for the existence of the various regimes shown in Table I and in Fig. (2). For instance, the question arises what plausible mechanism causes overshooting to occur in the polymerised fraction. This occurs, as per Table I, under conditions where the initial mean degree of polymerisation of active material is large and the polymerised fraction is much smaller compared to its equilibrium value. In that case, assembly starts with very few but very long filaments. These filaments can efficiently tend to equilibrium by breaking, and in the process create new nucleation centers. The number of segments a filaments has broken into depends on two factors: (1) length of the polymer and (2) probability of breaking of bonds (which in principle is also a function of length of the polymer). If a large polymer has been broken into a number of nucleation centers greater than the equilibrium number of polymers, then each newly created nucleation center extends towards the equilibrium length. This in turn causes the amount of polymerised material to be more than the equilibrium value and hence leads to the disintegration of some of the polymers in the end to satisfy the law of mass action. This is what we indeed observe as an overshoot in the polymerised fraction, which first increases to a value greater than equilibrium value and then decays towards equilibrium. We conclude that the overshoot in $f$ is directly connected with polymers being able to fragment by random scission.

Fig. (2a) and (2b) shows that conditions where overshoot happens depend quite strongly on the values of $\gamma$, our kinetic parameter. This is shown more clearly in Fig. (3a), where we show results obtained by numerically integrating our dynamical equations for different values of the parameter $\gamma$. For large enough $\gamma$, any transients in $f$ disappear completely. This suggests $\gamma$ regulates how prevelent scission is in the kinetic pathways. To substantiate this interpretation, we calculate the polymer density, $\rho_a = f \phi / \bar{N}_a$, where the active degree of polymerisation, $\bar{N}_a$, is related to the renormalised active degree of polymerisation via the relation, $\na=K_a \bar{N}_a^2$. As the activation constant, $K_a$, and the total monomer concentration, $\phi$, are not explicit parameters in our theory, we rewrite $\rho_a = K f/\sqrt{\na}$, where $K \equiv \phi \sqrt{K_a}$. In Fig. (3b) we show the scaled polymer density, $\rho_a / K$. From Fig. (3b), we see that $\rho_a$ also overshoots along with $f$, supporting our interpretation of the scission dominated polymer fraction overshoot. Also notice a sharp increase of polymer density for $\gamma \gg 1$ at short times. We do not fully understand the physical mechanism giving rise to this shape increase in the polymer density for short times.

\begin{figure}[ht!b]
        \centering
        \begin{subfigure}
            \centering
            \hspace{-13pt} \includegraphics[width=3.3in]{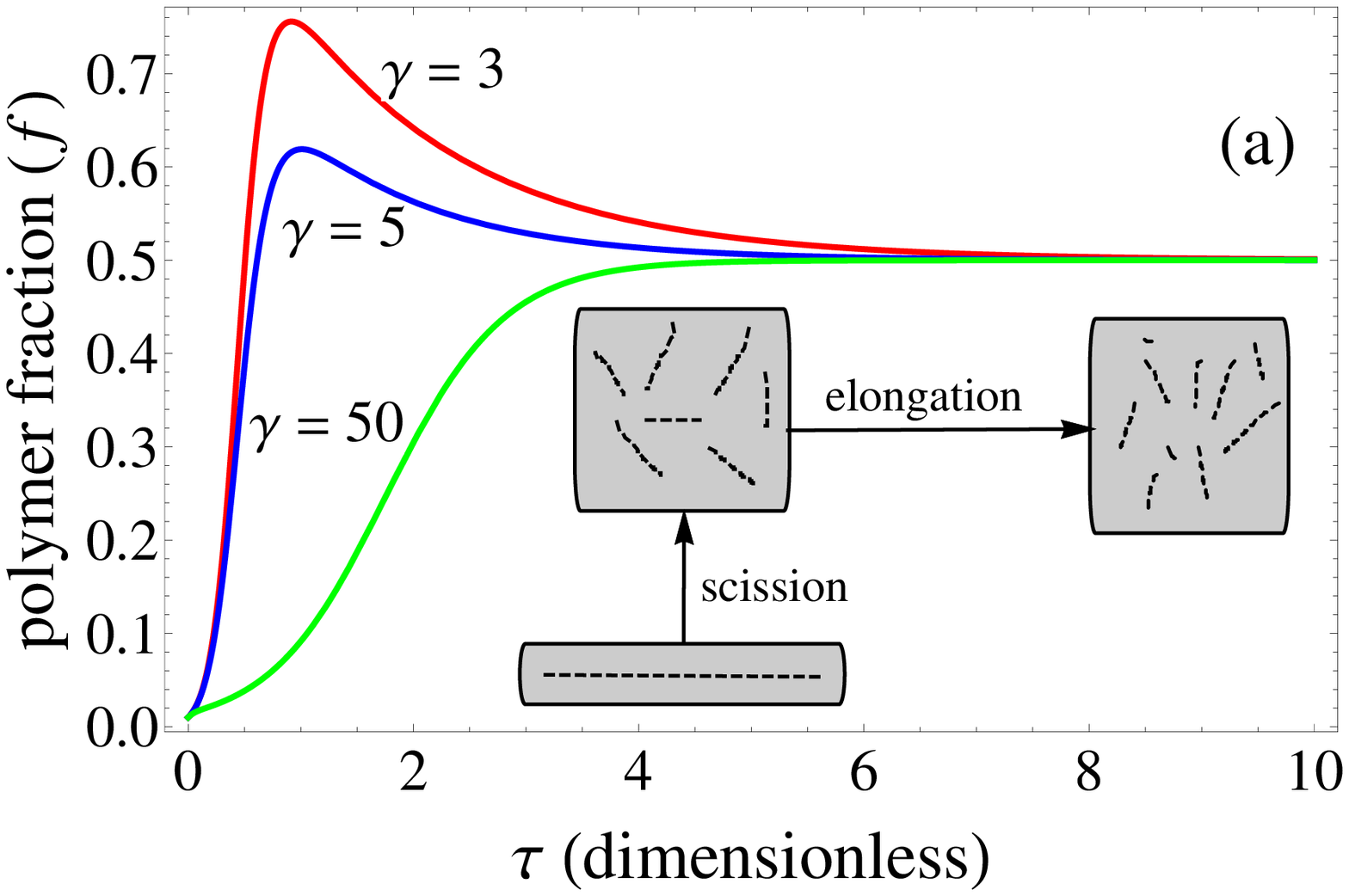}
        \end{subfigure}
        \hfill
        \begin{subfigure}
            \centering
            \includegraphics[width=3.3in]{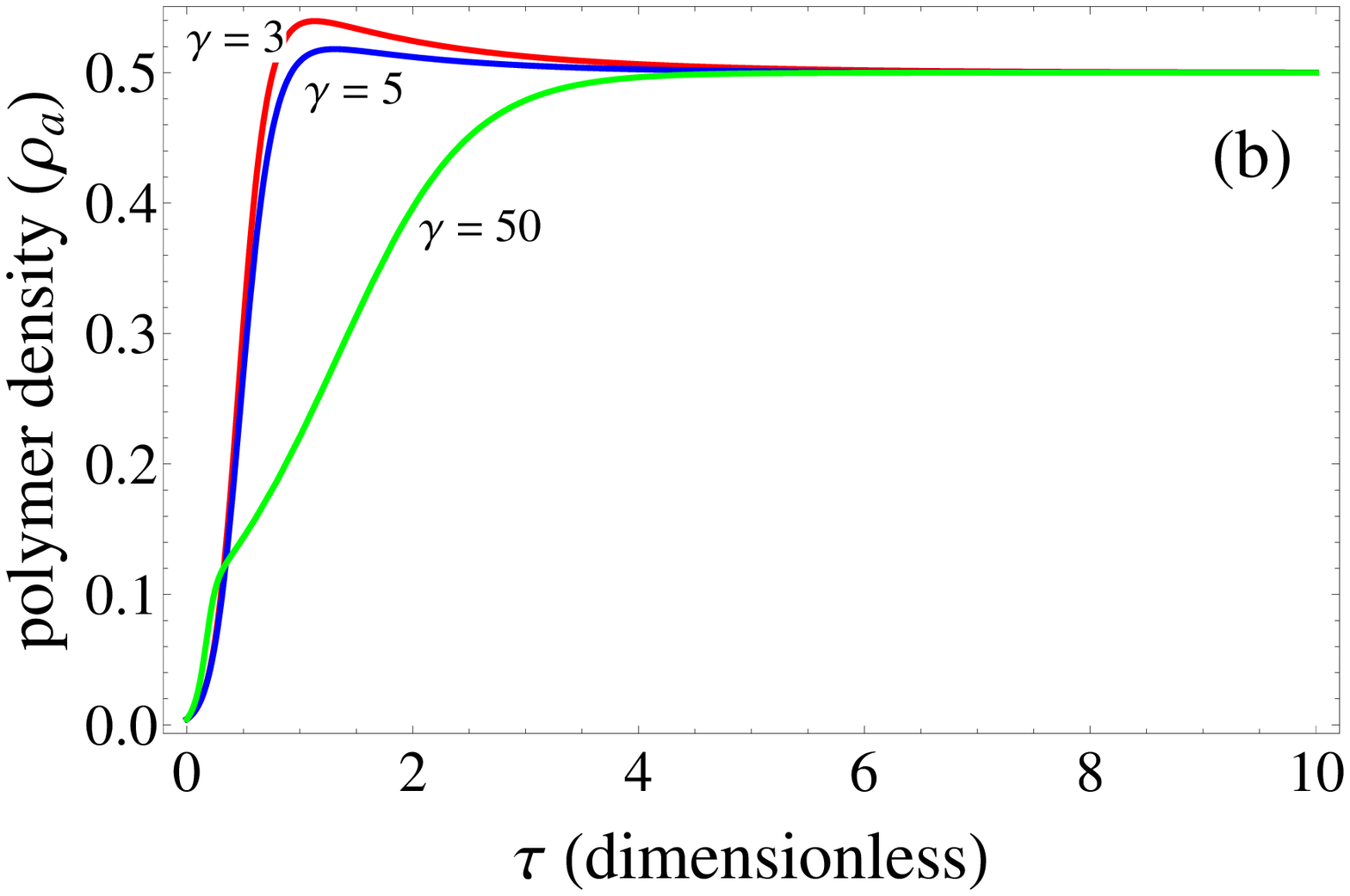}
            \begin{minipage}{.1cm}
            \vfill
            \end{minipage}
        \end{subfigure}
        \vspace{-15pt}
        \caption{a) Polymer fraction as a function of the dimensionless time, $\tau$, for values of our pathway controller $\gamma = 3,4,50$. $X=2$, $f(0)=0.01$ and $\na(0)=5$. The renormalised mean polymer length, $\na \equiv K_a \bar{N}_a^2$, for the same set of parameters shows monotonic response. Notice that increasing $\gamma$ suppresses overshoot, eventually suppressing it for $\gamma \ll 1$. For further explanation see the main text. b) Polymer density as a function of dimensionless time, $\tau$, for the same set of parameters as (a). See the main text for further explanation.}
\end{figure}

\section{Lag Time Analysis}
Sigmoidal response is a key characteristic of activated self-assembly and hence it should follow from our theory. Indeed, if we start off self-assembly with a very small initial fraction of polymerised material, we do find the sigmoidal behavior. This kind of response is characterized by a lag phase due to the time required to overcome the activation barrier, $\Delta f_a > 0$, and hence should become more pronounced as the height of this activation barrier increases or the initial polymer fraction, $f(0)$, decreases.
To analyse the lag time from our model, we use the conventional definition, that is, we find the time at which we have the maximum growth rate, determine the tangent at that point and identify its time intercept as the lag time.\cite{hellstrand} Of course, to be able to do this we have to supply the dynamical equations with initial conditions. The equilibrium relation between the polymerised fraction and the mean polymer length is $f = K_a \bar{N}_a (\bar{N}_a -1) /X$, but we arbitrarily choose $f(0) = K_a \bar{N}_a^2 (0)$ for $X=2$, and hence we start off with out-of-equilibrium initial conditions. It turns out that our results are qualitatively insensitive to the precise choise of initial conditions as long as we confine ourselves to the lower left part of the phase diagram Fig. (2).

In Fig. (4) we show by numerical solution that for increasing values of the initial polymerised fraction, $f(0)$, the lag time for assembly decreases and above a certain value the lag phase disappears completely. In fact, above a critical value we retrieve the transient behavior discussed in the previous section, see again Fig. (2). The origin of the lag time is the time required to create nucleation centers, which then grow into equilibrium polymers. Indeed, increasing the initial polymerised fraction is equivalent to seeding the system,\cite{cohen_review} and hence these pre-nucleation seeds then start elongating straight away, reducing the overall time required for nucleation.

\begin{figure}[!htb]
\begin{center}
\includegraphics[width=3.3in]{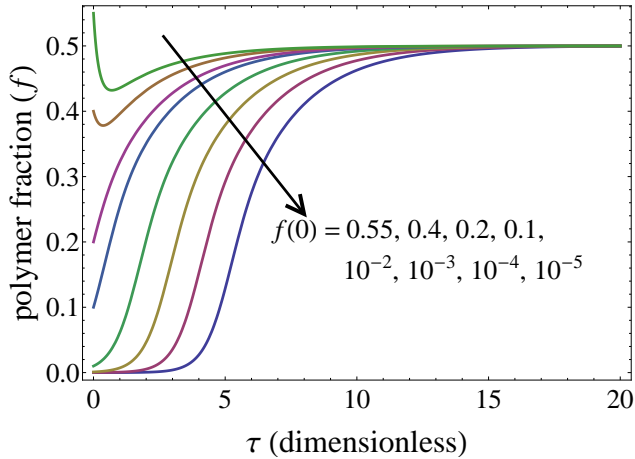}
\end{center}
\vspace{-15pt}
\caption{Polymer fraction, $f$, as a function of dimensionless time, $\tau$, for different values of the initial polymer fraction, $f(0)$. System parameters are $X=2$, $\gamma=2$ and $f(0)=\na(0)$.}
\end{figure}

From Fig. (4), which we obtained by means of numerical solution, we see that the lag time only occurs for initial polymerised fractions $f(0)$ close to zero, i.e., far away from the equilibrium value. If we want to obtain an analytical expression for the lag time, which will help us to compare with experimental data, we cannot rely on a linear analysis. Instead, we resort to perturbation theory.\cite{sidenote} Perturbation theory demands an expansion parameter around which an exact solution can be Taylor expanded. In principle, our phenomenological equations have two parameters, the mass action variable, $X$, and the pathway controller, $\gamma$. However, $\gamma$ seems to be the only sensible parameter to play the role of expansion parameter, as it is the only kinetic parameter.
It turns out that a lag time is very weakly dependent on the value of $\gamma$. To illustrate this, we solve our nonlinear equations numerically for fixed initial conditions and mass action variable, $X$, for different values of this parameter $\gamma$, and present results in Fig. (5). As Fig. (5) confirms, the lag time is relatively insensitive to $\gamma$, giving us a free choice of large or small $\gamma$ and still obtain the reliable estimate for the lag time. Notice that in Fig. (5), we have plotted the fraction of polymerised material as a function of the logarithm of time to see the behavior at large times, that is, to highlight the pseudo plateau that appears at intermediate times. The pseudo plateau eventually  equilibrates to true equilibrium, $f= 1-X^{-1}$, for large times. It appears that for small values of $\gamma$, our system of self-assembling monomers experiences a second lag phase, the origin of which eludes us. On the other hand, it points at two-stage nucleation seen in models where a disordered aggregate has to be nucleated before this in turn can nucleate in an ordered assembly which then can polymerize.\cite{knowles_two_stage} It shows again how rich the temporal behaviour of our model is.

After identifying $\gamma$ as our expansion parameter, Eqn. (8) and (9) give us the two options of $\gamma \ll 1$ and $\gamma \gg 1$. We notice that for the limiting case $\gamma \ll 1$, our system of differential equations is regular whilst for the opposite limiting case $\gamma \gg 1$ it is singular. This means that for the former we can put $\gamma = 0$ and hopefully get a convergent solution in powers of $\gamma$ by straightforward perturbation expansion. Because of its singular nature, for the latter  we cannot put $\gamma \rightarrow \infty$ and calculate perturbatively corrections in powers of $1/\gamma$. It turns out that, in spite of this, solving the singular version of our dynamical equations, i.e., for the case $\gamma \gg 1$, is much more simple. As from the numerical solutions we show that the short-time behavior of the system is different from the long-time behavior. Regular perturbation expansion breaks down for this kind of behavior whereas the technique known as the \textit{Matched Asymptotic Expansion} takes care of this.\cite{hinch} Matched Asymptotic Expansion has proven a useful scheme to provide reliable asymptotic solutions but only applies to singular problems. Hence, we choose to do our perturbation theory in the limit $\gamma \gg 1$, yet we expect accurate results for our lag time for all values of $\gamma$. Of course, our analysis for the behaviour at late times is only accurate for large $\gamma$ and completely misses the pseudo plateau shown in Fig. (5). We return to this issue below.

\begin{figure}[htb!]
\begin{center}
\includegraphics[width=3.4in]{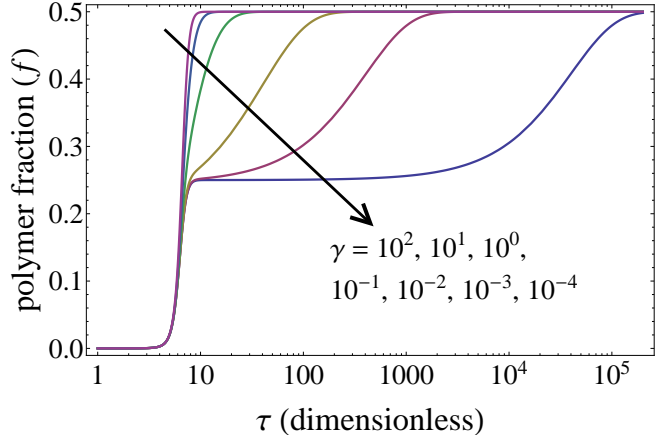}
\end{center}
\vspace{-15pt}
\caption{Polymerized fraction, $f$, as a function of logarithm of dimensionless time, $\tau$, for $X=2$, $f(0)= \na(0) = 10^{-6}$, showing that the lag time is extremely weakly dependent on $\gamma$. Also notice the pseudo plateau post lag-time for small values of our pathway controller $\gamma$. Also see the main text for discussion.}
\end{figure}

Invoking the method of Matched Asymptotic Expansion described in the appendix, we find from Eqn. (17) and (18), the following leading order solutions for $f(\tau)$ and $\na(\tau)$,
\bea
f(\tau) &=& \frac{(1-X^{-1})}{1+ \left( \frac{1-X^{-1}-f(0)}{f(0)} \right) e^{-4(1-X^{-1}) \tau}},
\eea
and
\bea
\na(\tau) &=& X f(\tau) - e^{- \frac{\gamma}{X} \tau} (f(0) X - \na (0)),
\eea
valid for $X >1$ and $f(0) \ll 1-X^{-1}$. Eqn. (18) is evidently a sigmoidal (logistic) function, which in fact we expect from Eqn. (8) and (9) taking the limit $\gamma \rightarrow \infty$. As the function form that we obtain for $f(t)$ is sigmoidal, it does not exhibit the transient phenomena of overshoot or undershoot. Notice that Eqn. (18) indeed is independent of $\gamma$ as we found from the numerical results shown in Fig. (5). In the limit $\gamma \rightarrow \infty$ we obtain the equality $f(\tau)= \na/X = K_a N_a^2(\tau) /X$ valid for all time, which in fact is the equilibrium condition for the fraction of polymerised material and the degree of polymerisation of the active material. The other equilibrium condition, $f=1-X^{-1}$ for $X>1$, is only reached for infinite time. Also, taking the limit $X \rightarrow \infty$ in Eqn. (18) and (19), we find that the sigmoidal response is preserved. This will turn out to have important consequences that we return to below.

As is done generally to analyse experimental assembly data, we can cast our sigmoidal relation in to following form and define an effective growth rate and lag time \cite{hellstrand}
\bea
f(\tau) = \frac{A}{1+e^{-k_{app} (\tau - \tau_{1/2})}},
\eea
where from Eqn. (18) we read off that $A = 1-X^{-1}$ is the equilibrium value, i.e., the saturation value, $k_{app} = 4 (1-X^{-1})$ is the effective growth rate and $\tau_{1/2} = (1-X^{-1})^{-1} \ln\left[\left( 1-X^{-1}-f(0)\right)f(0)^{-1}\right]$ is the time at which the maximum growth rate occurs, which turns out to be at the halfway point of assembly. Note that for $X \rightarrow 1$ the growth rate goes to zero and the half time, $\tau_{1/2}$, diverges, signifying what one may call critical slowing down by analogy to what happens in phase transitions near the critical point.\cite{ahart}

From Eqn. (17) and (19) we can now simply calculate the lag time, $\tau_{\text{lag}}$, by the method advertised above, i.e., by i) finding the time at which we have the maximum growth rate, ii) determining the tangent at that point, and iii) identifying the time intercept as the lag time. We find
\bea
\tau_{\text{lag}} = \frac{1}{4 (1- X^{-1})} \left[ \ln \left( \frac{1-X^{-1}-f(0)}{f(0)} \right) - 2 \right].
\eea
Eqn. (21), being the lag time calculated from the leading order solution for the polymerised fraction, $f$, is independent of $\gamma$. However, $\na$ does weakly depend on our pathway controller $\gamma$. As a consequence, this quantity can exhibit overshooting and undershooting in the large $\gamma$ limit. Note that strictly speaking our asymptotic solution is accurate in the limit $\gamma \gg 1$ implying that $\na$ is enslaved by $f$ and hence has the same lag time. The conditions for obtaining the transient response in the renormalised mean polymer length, $\na$, have been evaluated in the previous section.

To verify that the lag time is weakly dependent on the pathway controller, as we deduced from Fig. (5), we compare in Fig. (6) $\tau_{\text{lag}}$ for the polymerised fraction, $f$, from our analytical solution, Eqn. (20), and the results from a numerical solution of the governing equations. As the figure confirms, this turns out to be justified, indicating that  our asymptotic solutions for the analysis of the lag time are correct. This in turn implies that Eqn. (21) is a good estimate for any value of the pathway controller and determines under what conditions the lag phase exists. Eqn. (21) is real and positive only provided $f(0) < 1-X^{-1} = f(\infty)$, otherwise we lose the lag phase. This is particularly highlighted if we let $X \rightarrow 1^{+}$, where the lag phase vanishes unless $f(0) \rightarrow 0$ more quickly than $X$ goes to unity.

Eqn. (21) and Fig. (6) point at a remarkable property of our lag time, which is that it does \textit{not} vanish in the limit of an infinitely deep quench corresponding to taking the limit $X \rightarrow \infty$. This is remarkable, because naively one would expect that if the thermodynamic driving force becomes very large, this suppresses any phenomenon associated with nucleation. In our phenomenological theory the lag phase is the only response associated with nucleation and hence we expect it to vanish. Instead, in the limit $X \rightarrow \infty$, we find $\tau_{\text{lag}} \rightarrow \ln\left((1-f(0))/f(0)\right)/4-1/2 \equiv \tau_{\text{off}}$, i.e., the lag time tends to a finite non-zero value that we call the ``off time'', $\tau_{\text{off}}$.

\begin{figure}[htb!]
\begin{center}
\includegraphics[width=3.3in]{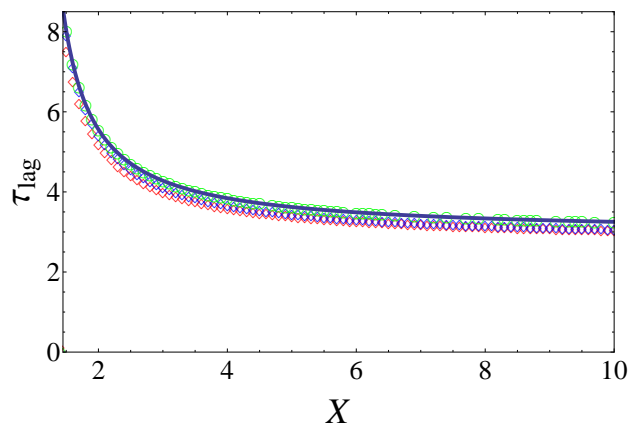}
\end{center}
\vspace{-15pt}
\caption{The dimensionless lag time, $\tau_{\text{lag}}$, as a function of the mass action variable, $X$, for values of the pathway controller $\gamma = 0.05 $ (red diamonds), $\gamma = 5 $ (blue diamonds) and $\gamma = 500 $ (green open circles), and initial conditions for the polymerised fraction, $f$, and the renormalised mean polymer length, $\na$, of $f(0) = \na(0) = 10^{-6}$. See also the main text. Symbols are numerical results while the continuous line is our analytical solution. }
\end{figure}

\section{Discussion}
We have seen in the previous section that depending, e.g., on the initial conditions, we can have a lag phase before assembly takes off. What we have not shown, however, is that in disassembly a lag phase is always absent. Hence, there is an inherent asymmetry between the polymerisation and depolymerisation kinetics. We call this asymmetry in the kinetics \textit{temporal hysteresis}. It is of interest to study in more detail in what way assembly is different from disassembly and what the parameters are that affect this. We confine ourselves to the hysteresis under conditions characterized by a lag phase in the assembly and use analytical solution for polymerised fraction, Eqn. (18), for that purpose.

To be able to compare the assembly and disassembly kinetics in a quantitative fashion, we perform a theoretical cyclic quench experiment, using our model. For this, we quench the system from one equilibrium state $X_1$ to another equilibrium state $X_2>X_1$, and after equilibration has taken place we quench back from $X_2$ to $X_1<X_2$. The former is an assembly process that potentially experiences a lag phase and the latter a disassembly process that does not have a lag phase. For a meaningful comparison between assembly and disassembly, we define the quantity $\Delta f(\tau) = \mid f(\infty) - f(\tau) \mid$, where $f(\infty)$ is the long-time (equilibrium) value of the polymerised fraction, $f$, after the quenching.
If assembly and disassembly do not exhibit hysteresis, the function $\Delta f(\tau)$ is identical for the two processes in our cyclic quench experiment. If there is hysteresis, this is no longer the case. We find that the magnitude of temporal hysteresis, to be specified in more detail below, depends on two factors: i) the width of the quench interval, $\Delta X = X_2 - X_1$, and ii) the distance of the quench interval from the critical point $X=1$. As to be expected, in the limit $\Delta X \rightarrow 0$, we do not find any hysteresis as the quench experiment is perturbative and hence the problem becomes a linear one. Also, as $X_1$ moves away from the critical point at $X=1$, the level of hysteresis decreases and, indeed, in the limit $X_1 \rightarrow \infty$ hysteresis is absent. The reason is that, as we have seen in the preceding sections, the lag phase for the assembly ceases to exist for the large initial polymerised fraction. In this case $\Delta X$ need not be very small for the hysteresis to vanish.

To support these two statements, we define a quantity that we call the hysteretic area, $A_h$, defined as the area between the kinetic curves, $\Delta f(\tau)$, for the assembly and disassembly in a cyclic quench experiment up to the point of intersection of the two, see also Fig. (7). To understand the reason for the existence of the hysteretic area, we revisit Eqn. (18), and note that the polymerised fraction, $f$, has two time scales: one associated with the apparent growth rate $k_{app}$ independent of initial conditions, and the half time $\tau_{1/2}$ that does depend on the initial polymerised fraction. The latter quantity becomes negative if the initial conditions are such that the system evolves to equilibrium via disassembly. As a consequence, disassembly evolves with only one of these two relaxation times, whereas assembly has both of them. This difference in time scales between assembly and disassembly gives rise to non-overlapping kinetic curves, and hence to a non-zero hysteretic area and also an intersection point. For this reason, it makes sense to focus on area between the two curves upto the point of intersection that gives us the quantitative measure for the level of hysteresis.

This is illustrated in Fig. (8), showing the hysteretic area, $A_h$, for various quenches between $X_1$ and $X_2$. We find that if we keep $X_1$ at some constant value but increase the quench interval, $\Delta X = X_2 -X_1$, by increasing $X_2$, the hysteretic area increases as a function of $\Delta X$. The figure also shows that the hysteretic area decreases as $X_1$ moves away from the critical point, $X=1$. This confirms that we can indeed conclude that the hysteretic response of polymerisation \textit{vs} depolymerisation disappears as we move away from critical point $X=1$ and hence is a characteristic of nucleated or activated self assembly.

\begin{figure*}[ht!b]
\begin{center}
\includegraphics[width=7in]{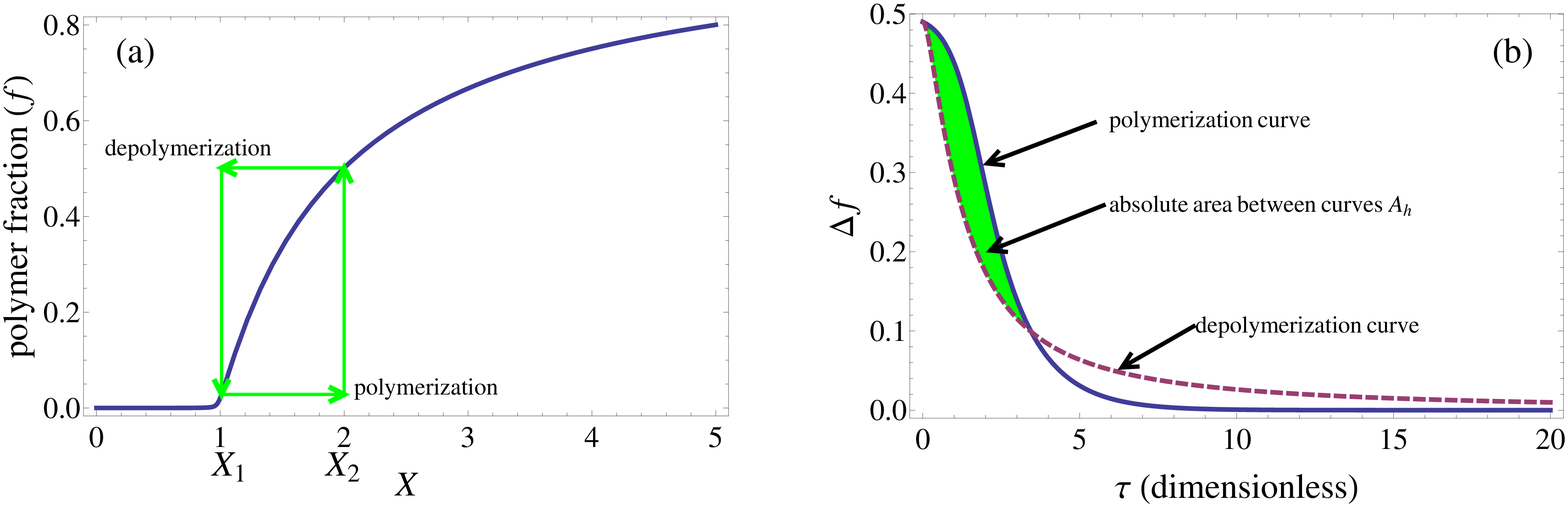}
\end{center}
\vspace{-15pt}
\caption{a) Mass action curve for polymer fraction, $f$ as a function of dimensionless mass action variable, $X$. Arrow show the quench cycle from $X_1$ to $X_2 > X_1$ (polymerisation) and vice versa (depolymerisation). b)Time dependence of the quantity $\Delta f(\tau) = \mid f(\infty) - f(\tau) \mid$, for the polymerisation (blue dashed curve) and depolymerisation (pink curve) dynamics when we quench system from one equilibrium ($X_1 = 1.01$) to an other equilibrium ($X_2 = 1.5$) point and back, for the pathway control variable, $\gamma =2$ and an activation constant,$K_a=10^{-5}$.}
\end{figure*}

\begin{figure}[ht!b]
\begin{center}
\includegraphics[width=3.3in]{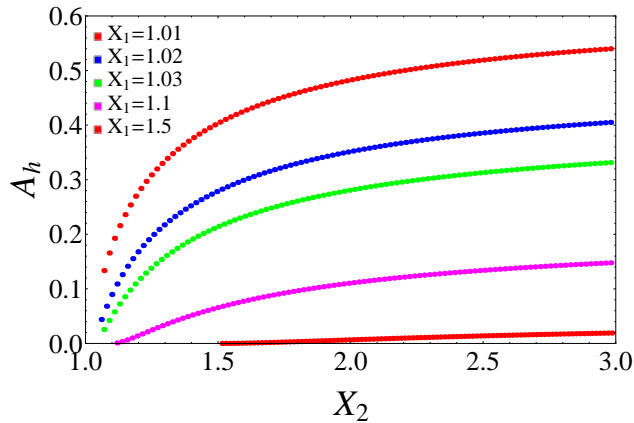}
\end{center}
\vspace{-15pt}
\caption{Absolute area between polymerisation and depolymerisation curves, $A_h$, obtained from analytical solutions for polymerised fraction, $f$, Eqn. (17), for quench interval $\Delta X = X_2-X_1$, where $X_1=1.01, 1,02, 1.03, 1.5$  are kept constant but $\Delta X$ increases as a consequence of an increasing $X_2$. See also the main text.}
\end{figure}

We used our results of section (IV) for the lag phase to study the potential kinetic asymmetry between the assembly and disassembly. Our analysis of the lag phase along with that of the transient response, discussed in section (III), turns out to allow us to provide a physical or molecular interpretation of the phenomenological parameters in our theory. Our phenomenological model has two important parameters: the thermodynamic mass action variable, $X$, and the phenomenological kinetic parameter, $\gamma$. $X$ describes the final thermodynamic state of the solution, as we have shown in section (II), whereas $\gamma$ controls the temporal evolution of the polymerised fraction, $f$, and the mean polymer length of active material, $\bar{N}_a$, towards that final thermodynamic state that potentially takes place via a plethora of pathways.\cite{cates5}

The relative weight of each of the pathways governs in the end the precise kinetics of our linearly polymerizing system, and this suggests that there must be a connection between the parameter $\gamma$ and the relative weight of these pathways. This is why we referred to it as the pathway controller. To justify that $\gamma$ must indeed be a pathway controller, we revisit Fig. (5) showing how the parameter $\gamma$ influences in what way the polymerised fraction evolves as a function of time. Initial conditions are such that at zero time the fraction of polymerised material is vanishingly small and (eventually) equilibrates to the value of one half. In the limit $\gamma \rightarrow 0$, however, we obtain a pseudo plateau  that arguably hints at how the pathways and the value of $\gamma$ are related. This pseudo plateau emerges because of a diverging time scale that emerges post lag-phase, in the late-stages of the assembly process. We also find a similar pseudo plateau for the mean polymer length, $\bar{N}_a$.

The reason that we associate this psuedo plateau with a specific reaction pathway, is that Semenov and Nyrkova find within the rate equation approach a similar late-stage diverging time scale for the so-called scission-recombination pathway.\cite{semenov2} They conclude that in the limit $K_a \rightarrow 0$, the mean polymer length of the active material displays a signature of critical slowing down in particular for $X \rightarrow 1^+$. However, the critical region vanishes in that same limit $K_a \rightarrow 0$ and hence should not be observable in the limit where our theory is valid. What remains in the Semenov-Nyrkova theory is the time scale that scales as $K_a^{-2/3}$ and that time scale diverges irrespective of the value of the mass action variable, $X$, which is also true for our model.\cite{semenov2} So, this suggests that our $\gamma \rightarrow 0$ limit should represent reversible scission-recombination pathway.
Other aspects of the kinetics point in the same direction, in particular the transients. Indeed, we observe in Fig. 3 that the magnitude of the overshoot increases with decreasing value of the pathway controller, $\gamma$. In section (III), we rationalized the existence of an overshoot in terms of almost instantaneous scission of very few, very long polymers combined with a somewhat slower unspecified growth mechanism. As $\gamma$ increases, overshooting becomes less prominent and disappears completely for $\gamma \gg 1$: this ties in with a decreasing ability of the polymers to break and create nucleation centers that then can grow.  As $\gamma \rightarrow \infty$, scission must be absent within our interpretation. The polymers will in that case be nucleated from inactive monomers and hence we see the lag phase emerging instead of an overshoot.

In conclusion, we have a kinetic parameter in our phenomenological theory that plausibly acts as a pathway controller. A natural question that now arises is: if $\gamma$ is indeed the pathway controller, then why is the lag time independent of it? To answer this question, we need to understand that the lag time arises as a consequence of the existence of a nucleation barrier and effectively takes place at the monomer level, i.e., does not involve any polymers. The reason is that an assembly inactive monomer needs to transition to an assembly active state before it can partake in a polymerisation reaction. Naively, the lag phase is controlled on the one hand by the free energy difference between the active and inactive monomer states, and on the other by the thermodynamic driving force towards the polymerised state. This explains why, as $X \rightarrow 1$, the lag time diverges as the thermodynamic driving force becomes zero. It does not explain, however, why in our model, even for infinite thermodynamic driving force, so for $X \rightarrow \infty$, the lag time does not vanish.

Let us now confront our predictions for the lag time with actual experimental data on Amyloid-$\beta$ assembly. (Because of a lack of data we cannot do the same for the transient response and hysteresis.) Hellstrand \textit{et al.} obtained lag times as a function of the concentration of Amyloid$\beta$(M1-42) protein in aqueous solution at pH $=8$, containing 20 mM sodium phosphate, 200$\mu$M EDTA, 0.02$\%$ $\text{NaN}_3$, and $20\mu$M ThT.\cite{hellstrand} The experimental data together with our theoretical fits are shown in Fig. (9).
To obtain our theoretical fits, we need to fix our mass action variable, $X$, the initial value for the fraction of polymerised material, $f(0)$, and the relaxation rate $\Gamma_1$ as defined in section (II). The mass action variable, $X=\phi \exp(-\Delta f_e / k_B T)$, we determine by realizing that the experiments are done at constant physico-chemical conditions, and hence that the elongation constant, equal to $\exp(-\Delta f_e / k_B T)$ is fixed. Because the critical point is at $X=1$, we deduce that the elongation constant must be equal to the reciprocal of the critical concentration, $\phi_c$. Hence, $X=\phi / \phi_c = C/C_c$ where $C$ and $C_c$ are now the dimension bearing molar concentrations. The critical concentration $C_c \sim 0.18\mu\text{M}$ we determine from the experimentally determined concentration below which no polymers are is detected.\cite{hellstrand}
The corresponding value of the dimension-bearing elongation constant turns out to be $5.6\times 10^{6}  \text{M}^{-1}$, which corresponds to a binding free energy $\Delta f_e$ of about $-$20 $k_B T$.

Our lag time expression Eqn. (21) depends logarithmically on the initial polymerised fraction, $f(0)$. From the experiments of Hellstrand et al., we do not know its value. Hence, we choose $f(0)$ as well as our phenomenological kinetic parameter $\Gamma_1$ in such a way to get the best agreement for low and high concentrations. We recall that $\Gamma_1$ is the fundamental relaxation rate for the polymerised fraction in our model, see Eqn. (5). As we have discussed in previous section, our lag time remains non-zero even for infinite concentration, whereas in the experiments the lag time does tend to zero with increasing concentration. Hence, to fit our theoretical results to the experimental data we need to subtract the off time $\tau_{\text{off}}=\ln\left((1-f(0))/f(0)\right)/4-1/2$ that we defined in the preceding section and that depends only on the initial condition.
We get reasonable agreement with the experiments if we set $f(0)=10^{-6}$ and $\Gamma_1=5.21 \text{1/hour}$, telling us that a theory captures in essence the concentration dependence of the lag time of this particular system.

Clearly, the fact that we have to subtract a known off time is unsatisfactory, although this does not preclude the possibility of systems that do show such a response. Mathematically, the off time in our theory is caused by the specific form of the free energy that we constructed. This leads to kinetic equations that remain of a logistic form even in the limit, $X \rightarrow \infty$. We have not been able to reformulate the theory in such a way that it suppresses the existence of an off time. The advantage of the model is that it is simple, that it exhibits a very rich kinetic behavior and that it can relatively straightforwardly be extended to include other macroscopic phase transitions. In future work, we connect this theory to the Landau-de Gennes theory for the isotropic-nematic phase transition, allowing us to model phase ordering kinetics in solutions of chromonics or surfactants.

\begin{figure}[ht!b]
\begin{center}
\includegraphics[width=3.3in]{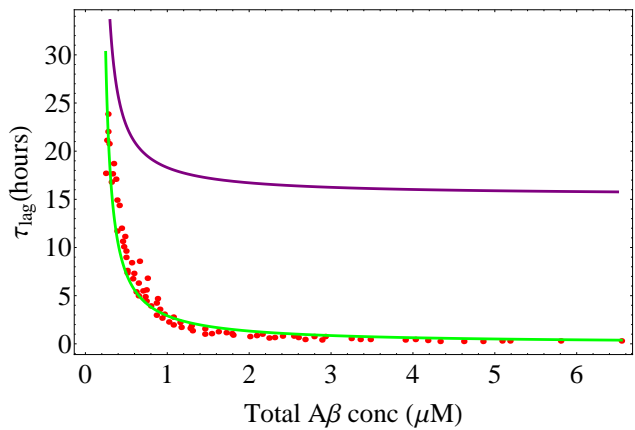}
\end{center}
\vspace{-15pt}
\caption{Lagtime as a function of $\phi$, Dots are experimental results from Erik Hellstrand et. al. while continuous line is analytical solution, Fit parameters are $\Gamma_1=5.21$, $K_e=5.8 \times 10^6$ and $f(0)=10^{-6}$, purple curve is $\tau_{\text{lag}}$ asymptotic solution and green curve is $\tau_{\text{lag}}-\tau_{\text{off}}$}
\end{figure}

\section{Conclusion}
In this work we present a phenomenological Landau theory for nucleated linear self assembly. Our model is consistent with the thermodynamics of linear polymerisation in the limit $K_a \rightarrow 0$, i.e., for small values of the activation constant, and is able to produce all transient behaviours observed experimentally, i.e., a lag phase to assembly as well as overshoot and undershoot. We show that the transient response of overshoot and undershoot is not caused by the nonlinearity of the problem in hand but instead a consequence of the competition between the polymerised fraction, $f$, and the mean polymer length, $\bar{N}_a$. Both the types of transient behaviour can be obtained from the linearised version of our theory. We pinpoint the initial conditions required to observe overshoot or undershoot in the polymerised fraction, $f$ and that in the mean polymer length, $\bar{N}_a$.

Solving our nonlinear dynamical equations for the limit $\gamma \rightarrow \infty$ using the method of Matched Asymptotic Expansion, we obtain an analytical expression for the lag time, $\tau_{\text{lag}}$, which presents itself only when at time zero very little of the material is polymerised. The lag phase can only be found as a non-linear response, so does not follow from the linearised theory. By comparing numerical and analytical solutions we show that the lag time is for all intends and purposes independent of our phenomenological kinetic parameter $\gamma$ that we identified as the pathway controller. We find different kinetic profiles for polymerisation and depolymerisation, call this a temporal hysteresis, and show that this temporal hysteresis is due to the inherent asymmetry involved in the polymerisation and depolymerisation. The reason, of course is that a nucleation barrier needs to be crossed only upon polymerisation. When we move deeply into the polymerised regime, hysteresis becomes less prominent.

By comparing the temporal behavior of the polymerised fraction in the limit $\gamma \rightarrow 0$ with microscopic theory for the scission and recombination kinetic pathway,\cite{semenov2} we argue that in this limit our model is dominated by scission and recombination. We find that the magnitude of transient overshoot and undershoot increases with decreasing value of $\gamma$. This finding, explains why $\gamma$ indeed must regulate the predominance of scission kinetics.

If we compare our theoretical lag time with experimental data on Amyloid-$\beta$, our theory agrees well with the experiments, apart from an additive constant that we need to remove. The additive constant that we call the off time, $\tau_{\text{off}}$, turns out to be an asymptotic value of the lag time for very large supersaturation. We do not fully understand the origin of this off time in our phenomenological theory. Finally, our phenomenological Landau theory, whilst much simpler than more conventional rate equation approaches to self-assembly, mimics many, if not all of the generic features, seen in theory and experiments, including hysteresis.

\section{Acknowledgement}
The authors would like to thank Tom de Greef (TU/e) and Laurent Bouteiller (UPMC) for useful discussions. This work was supported by the Nederlandse Organisatie voor Wetenschappelijk Onderzoek through Project No. 712.012.007.

\appendix
\section{Specific Heat Calculation}
Here, we calculate specific heat at constant volume, and show its agreement with the same quantity obtained from mass action theory. As in conventional Landau theory there is a linear relation between the Landau free energy and the temperature. Therefore, we start out with a linear approximation in the vicinity of the polymerisation point. Using the definition $X = \phi e^{-\Delta f_e /k_B T}$, in the vicinity of the polymerisation point at $X =1$, this quantity can be expressed as,
\bea
X = \phi e^{\Delta f_p / k_B T_p + \frac{1}{k_B T_p^2} \Delta h_p (T - T_p)}
\eea
by Taylor expanding $\Delta f_e / k_B T$ areound the polymerisation temperature $T_p$, where $\Delta f_p$ and $\Delta h_p$ are the free energy and enthalpy of binding at $T_p$. Evidently $X$ at the polymerisation point is given by
\bea
X_p = \phi e^{-\Delta f_p / k_B T_p} =1. \nonumber
\eea
This yields
\bea
X \approx e^{\frac{\Delta h_p}{k_B T_p^2} (T - T_p)} \approx 1+ \frac{\Delta h_P}{k_B T_p^2} (T - T_p),
\eea
provided $|\Delta h_p (T - T_p) / k_B T_p^2| \ll 1$. If the Landau free energy density is considered, it follows from Eqn. (4) that in equilibrium $S_1^2 = S_2$, and therefore the free-energy density can be expressed as, so inserting (A2) then gives
\bea
\frac{F}{k_B T_p} \equiv - \frac{\Delta h_p}{k_B T_p^2} (T - T_p) S_1^2 + \frac{1}{2} S_1^4 .
\eea
Note that the free energy function is dimensionless, and we make a choice of thermal energy at the polymerisation temperature, i.e., $k_B T_p$ to non-dimensionalise our free energy.
In equilibrium
\bea
S_1 &=& 0 \vee S_1^2 = \frac{\Delta h_p}{k_B T_p^2} (T - T_p).
\eea
Note that for $h_p > 0$, the second equality of Eqn. (A4) holds for $T > T_p$, whilst if $h_p < 0$, it holds for $T < T_p$.

We arbitrarily assume that $h_p < 0$, and the free-energy density can now be computed for $T \le T_p$,
\bea
\frac{F}{k_B T_p} = - \frac{1}{2} \left( \frac{\Delta h_p}{k_B T_p^2} \right)^2 (T - T_p)^2,
\eea
where $F=0$ for $T > T_p$.

Hence, the isochoric heat capacity per unit of particle $c_v$ can then be computed from,
\bea
c_v = -T \frac{\partial^2 F}{\partial T^2} = T \frac{\Delta h_p^2}{k_B T_p^3},
\eea
for $T \le T_p$ and $c_v =0$ for $T \ge T_p$. Eqn. (A6) agrees with specific heat at the critical temperature, $T_p$, calculated by van Jaarsveld et al., after some algebra, for the so-called thermally activated polymerizing systems.\cite{paul2}

\section{$\gamma \gg 1$ solution using Matched Asymptotic Expansion}
In this appendix we outline the method of Matched Asymptotic Expansion,\cite{hinch} by virtue of which we obtained analytical solutions for our dynamical equations, (8) and (9), in the limit $\gamma \rightarrow \infty$. The method of matched asymptotic expansion, principally involves four steps: 1) obtaining the outer solution for long times $\tau \gg 1/\gamma$, 2) finding the inner solution for short times $\tau \ll 1/\gamma$, 3) matching the inner and outer solutions in the intermediate time domain and 4) obtaining a composite solution valid for the whole time domain.

Starting with the outer solution, we first assume that the solution has a regular Taylor expansion for $\epsilon$,
\bea
f^{out}(\tau,\epsilon) &=& f_0^{out}(\tau) + \epsilon f_1^{out}(\tau) + O(\epsilon^2),
\eea
and
\bea
\na^{out}(\tau,\epsilon) &=& \nazero^{out}(\tau) + \epsilon \naone^{out}(\tau) + O(\epsilon^2).
\eea
Substituting this into Eqn. (8), we obtain the zeroth-order equations,
\bea
\frac{df_0^{out}(\tau)}{d \tau} &=& 4 (1 - X^{-1}) f_0^{out}(\tau) + 4 X^{-1} f_0^{out}(\tau) \nazero^{out} (\tau) \nonumber \\
 & & - 8 f_0^{out}(\tau)^2,
\eea
and
\bea
0 &=& X  f_0^{out}(\tau) - \nazero^{out} (\tau),
\eea
where the subscript ``0'' refers to the order of the solutions.

We cannot impose both initial conditions on the leading-order outer solutions. We therefore take most general solution of these equations. As we shall see, when we come to matching that to the inner solution, the natural choice of
imposing the initial condition $f_0^{out} (0) = f(0)$ is in fact correct. From the equation (B4), we conclude that
\bea
\nazero^{out} (\tau) = X f_0^{out}(\tau), \nonumber
\eea
for all $\tau \ge 0$. The zeroth order degree of polymerisation $\nazero^{out}$ corresponds to quasi-equilibrium for the fraction of active material $f_0^{out}$, in which the degree of polymerisation increases because of an increase in active material. The degree of polymerisation decreases also because of breaking of long filaments, hence increasing number of polymers.

Substituting this result into equation (B3), we get a first order differential equaiton for $f_0(\tau)$,
\bea
\frac{df_0^{out}(\tau)}{d \tau} = 4 (1 - X^{-1}) f_0^{out}(\tau) - 4 f_0^{out}(\tau)^2.
\eea
The solution of this equation is given by
\bea
f_0^{out} (\tau) &=& \frac{4 (1 - X^{-1}) \alpha e^{4 (1 - X^{-1}) \tau}}{4 (1 - X^{-1}) - 4 \alpha + 4 \alpha e^{4 (1 - X^{-1}) \tau}},
\eea
and hence
\bea
\nazero^{out} (\tau) &=& \frac{4 (X -1) \alpha e^{4 (1 - X^{-1}) \tau}}{4 (1 - X^{-1}) - 4 \alpha + 4 \alpha e^{4 (1 - X^{-1}) \tau}}.
\eea
Here $\alpha$ is a constant of integration. This solution is invalid near $\tau =0$, because no choice of $\alpha$ can satisfy the initial conditions for both $f_0^{out}$ and $\nazero^{out}$.

To solve the problem for short times, we surmise that there is a short initial layer, for times $t = O (\epsilon)$, in which $f$ and $\na$ adjust from their initial values to values that are compatible with the outer solution found above. We introduce the inner variables
$T = \tau/\epsilon$, $f^{in}(T, \epsilon) = f^{out}(\tau, \epsilon)$
and $\na^{in}(T, \epsilon) = \na^{out}(\tau,\epsilon)$.
The inner equations then become,
\bea
\frac{d f^{in}(T,\epsilon)}{dT} &=& \epsilon ( 4 (1-X^{-1}) f^{in}(T, \epsilon), \\
&+& 4 X^{-1} f^{in}(T, \epsilon) \na^{in} (T, \epsilon) - 8 f^{in} (T, \epsilon)^2 ), \nonumber
\eea
and
\bea
\frac{\na^{in} (T, \epsilon)}{d T} &=& X f^{in} (T, \epsilon) - \na^{in} (T, \epsilon),
\eea
with boundary conditions, $f^{in} (0,\epsilon) = f(0)$ and $\na^{in} (0,\epsilon) = \na(0)$.

We look for an inner expansion of the form
\bea
f^{in}(\tau,\epsilon) &=& f_0^{in}(\tau) + \epsilon f_1^{in}(\tau) + O(\epsilon^2),
\eea
and
\bea
\na^{in}(\tau,\epsilon) &=& \nazero^{in}(\tau) + \epsilon \naone^{in}(\tau) + O(\epsilon^2).
\eea
The leading order inner equations then become
\bea
\frac{d f_0^{in}}{dT} &=& 0,
\eea
and
\bea
\frac{d \nazero^{in}}{dT} &=& X f_0^{in} - \nazero^{in},
\eea
with boundary conditions,
$f_0^{in} (0,\epsilon) = f(0)$ and $\nazero^{in} (0,\epsilon) = \na(0)$.

The solution of Eqn. (B12) and (B13) are
\bea
f_0^{in} (T) &=& f(0),
\eea
and
\bea
\nazero^{in} (T) &=& e^{-T} \left( - f(0) X + e^T f(0) X + \na(0) \right).
\eea
Now that we have obtained expressions for the inner and outer solution, we assume both valid for intermediate times of the order $\epsilon \ll \tau \ll 1$. We require that the expansions agree asymptotically in this regime, where $T \rightarrow \infty$ and $\tau \rightarrow 0$ as $\epsilon \rightarrow 0$. Hence, the matching conditions must read, $\lim_{T \rightarrow \infty} f_0^{in} (T) = \lim_{\tau \rightarrow 0^+} f_0^{out} (\tau) = f(0)$ and $\lim_{T \rightarrow \infty} \nazero^{in} (T) = \lim_{\tau \rightarrow 0^+} \nazero^{out} (\tau) = f(0) X$.

The condition implies that, $f_0^{out} (0) = \alpha = f(0)$ and $\nazero^{out} = f(0) X$.

This implies that the outer solutions become,
\bea
\hspace{-20pt} f_0^{out} (\tau) &=& \frac{4 (1 - X^{-1}) f(0) e^{4 (1 - X^{-1}) \tau}}{4 (1 - X^{-1}) - 4 f(0) + 4 f(0) e^{4 (1 - X^{-1}) \tau}},
\eea
and
\bea
\hspace{-20pt} \nazero^{out} (\tau) &=& \frac{4 (X -1) f(0) e^{4 (1 - X^{-1}) \tau}}{4 (1 - X^{-1}) - 4 f(0) + 4 f(0) e^{4 (1 - X^{-1}) \tau}}.
\eea
Having now obtained expressions for the first terms of both the inner and the outer expansions, they must now be matched together to obtain one composite expansion that approximates the solution over the whole time domain.
To get the composite expansion, the inner and outer expansions are simply added together and the common limit found in (B16) is subtracted, for otherwise it would be included twice in the overlapping region. So our composite solution
finally reads,

\bea
f &\sim& f_0^{out} (\tau) + f_0^{in} (\frac{\tau}{\epsilon}) - f(0),
\eea
for the fraction of active material
\bea
\na &\sim& \nazero^{out} (\tau) + \nazero^{in} (\frac{\tau}{\epsilon}) - f(0) X,
\eea
for the renormalised degree of polymerisation of the active material.

Keeping in mind that $\epsilon = 1/\gamma$, gives us the full zeroth order solution given in Eqn. (18) and (19).


\begin{thebibliography}{}
\bibitem{sidenote} One possible alternative would be recursive iteration that requires initial guess for an approximate solution. The advantage of such an approach is that it is free of any small parameter but the convergence of the iteration is highly dependent on the trial solution.\cite{cohen1, cohen2, cohen3} Unfortunately, we have  not been able to find a suitable trial solution that converges to an exact solution.
\end{thebibliography}
\end{document}